\documentclass[twocolumn,showpacs,preprintnumbers,amsmath,amssymb,superscriptaddress]{revtex4-1}
\usepackage[english]{babel}
\usepackage[dvips]{graphicx}
\usepackage{dcolumn}
\usepackage{bm}

\usepackage{color}

\begin{document}
\title{Dynamics of self-generated, large amplitude magnetic fields following \\high-intensity laser matter interaction}
\author{G. Sarri}
\affiliation{School of Mathematics and Physics, The Queen's University of Belfast, Belfast, BT7 1NN, UK}
\author{A. Macchi}
\affiliation{Istituto Nazionale di Ottica, Consiglio Nazionale delle Ricerche, research unit ``Adriano Gozzini'', Pisa, Italy}
\affiliation{Dipartimento di Fisica ``E. Fermi'', Largo B. Pontecorvo 3, I-56127 Pisa, Italy}
\author{C. A. Cecchetti}
\affiliation{Light4Tech s.r.l, 50018 Scandicci, Italy}
\author{S. Kar}
\affiliation{School of Mathematics and Physics, The Queen's University of Belfast, Belfast, BT7 1NN, UK}
\author{T. V. Liseykina}
\affiliation{Institut f\"{u}r Physik, Universit\"{a}t Rostock, D-18051 Rostock, Germany}
\author{X. H. Yang}
\affiliation{School of Mathematics and Physics, The Queen's University of Belfast, Belfast, BT7 1NN, UK}
\author{M. E. Dieckmann}
\affiliation{School of Mathematics and Physics, The Queen's University of Belfast, Belfast, BT7 1NN, UK}
\author{J. Fuchs}
\affiliation{LULI, Ecole Polytechnique, CNRS, CEA, UPMC; 91128 Palaiseau, France}
\author{M. Galimberti}
\affiliation{Rutherford Appleton Laboratory, Central Laser Facility, Chilton, OX11 0QX, UK}
\author{L. A. Gizzi}
\affiliation{Istituto Nazionale di Ottica, Consiglio Nazionale delle Ricerche, research unit ``Adriano Gozzini'', Pisa, Italy}
\author{R. Jung}
\affiliation{Institute for Laser and Plasma Physics, Heinrich Heine University, Dusseldorf, Germany}
\author{I. Kourakis}
\affiliation{School of Mathematics and Physics, The Queen's University of Belfast, Belfast, BT7 1NN, UK}
\author{J. Osterholz}
\affiliation{Institute for Laser and Plasma Physics, Heinrich Heine University, Dusseldorf, Germany}
\author{F. Pegoraro}
\affiliation{Istituto Nazionale di Ottica, Consiglio Nazionale delle Ricerche, research unit ``Adriano Gozzini'', Pisa, Italy}
\affiliation{Dipartimento di Fisica ``E. Fermi'', Largo B. Pontecorvo 3, I-56127 Pisa, Italy}
\author{A. P. L. Robinson}
\affiliation{Rutherford Appleton Laboratory, Central Laser Facility, Chilton, OX11 0QX, UK}
\author{L. Romagnani}
\affiliation{LULI, Ecole Polytechnique, CNRS, CEA, UPMC; 91128 Palaiseau, France}
\author{O. Willi}
\affiliation{Institute for Laser and Plasma Physics, Heinrich Heine University, Dusseldorf, Germany}
\author{M. Borghesi}
\affiliation{School of Mathematics and Physics, The Queen's University of Belfast, Belfast, BT7 1NN, UK}

\date{\today}

\begin{abstract}
The dynamics of magnetic fields with amplitude of several tens of Megagauss, generated at both sides of a solid target irradiated with a high intensity ($\sim10^{19}$W/cm$^2$) picosecond laser pulse, has been spatially and temporally resolved using a proton imaging technique. The amplitude of the magnetic fields is sufficiently large to have a constraining effect on the radial expansion of the plasma sheath at the target surfaces. These results, supported by numerical simulations and simple analytical modeling, may have implications for ion acceleration driven by the plasma sheath at the rear side of the target as well as for the laboratory study of self-collimated high-energy plasma jets.
\end{abstract}


\maketitle

The generation of magnetic fields in plasmas is a phenomenon of great relevance for a wide range of physical scenarios and it has been studied in laser-produced plasmas since the introduction of high-power lasers, with particular emphasis to their role in Inertial Confinement Fusion \cite{Stamper}. In this scenario, involving nanosecond laser pulses of intensity $I_L\approx 10^{14}- 10^{16}$ W/cm$^2$, magnetic field generation is generally described by hydrodynamic modeling \cite{Haines} and it has been recently characterized by temporally and spatially resolved measurements \cite{Cecchetti,Petrasso}. In the case of shorter ($\leq$ ps) and more intense ($I_L\geq 10^{18}$ W/cm$^2$) laser pulses, the production of large currents of high energy ``hot'' electrons requires a more complex modeling \cite{Mason} and very intense magnetic field may be generated by, e.g., Weibel-like instabilities \cite{Wagner} or electron recirculation at the plasma boundary ("fountain effect") \cite{fountain}. In this latter case, magnetic fields may influence the emission of multi-MeV proton \cite{Miyazaki,Pukhov} and positron \cite{Chen} beams.
Magnetic fields generated in relativistic laser-produced plasmas are also of central importance in reproducing conditions resembling large-scale astrophysical processes \cite{Remington} on a laboratory scale such as the self-collimation of relativistic leptonic jets \cite {Honda} or the upstream-downstream mixing in supernova remnant shocks \cite{Martins}.

Previous experimental work has detected effects induced by such magnetic fields on external optical beams \cite{Borghesi} or on the polarization of self-generated harmonics \cite{Tatarakis}. However, these measurements suffered of limitations, in terms either of the range of plasma density accessible \cite{Borghesi} or of spatial and temporal resolution \cite{Tatarakis}. Furthermore, only the fields generated at the front (laser-irradiated) side of the target have been investigated; the magnetic field generation at the rear side, where proton acceleration in the expanding fast electron sheath takes place \cite{Snavely}, is thus yet to be experimentally characterized.

In this Letter we report on simultaneous measurements of the magnetic fields generated at the front and rear side of a solid target irradiated by a short and intense laser pulse, using a spatially and temporally resolved proton imaging technique \cite{Sarri}. Clear evidence is given of the generation of toroidal magnetic fields (maximum amplitude of $\sim50$ MegaGauss) that decay in time on a picosecond time scale. Their spatial distribution and amplitude is consistent with the recirculation of the laser-accelerated electrons around the target and they are sufficiently intense to confine the radial plasma expansion.

\begin{figure*}[!t]
\begin{center}
\includegraphics[width=2\columnwidth]{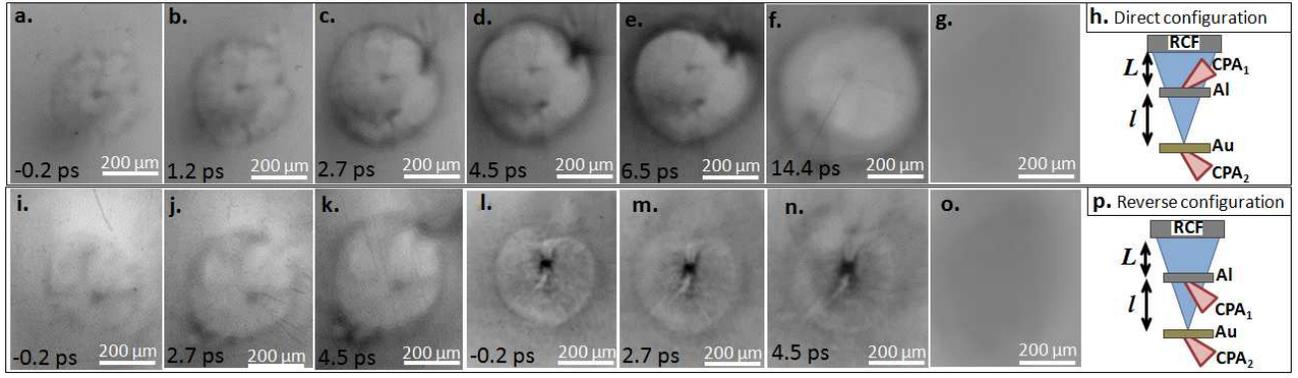}
\caption{\textbf{a.-f.} Proton imaging snapshots in the direct configuration (sketched in the frame \textbf{h.}) and typical image of the unperturbed proton beam (\textbf{g.}). \textbf{i.-k.} Additional set of images, still obtained in the direct configuration, from a different laser shot. \textbf{l.-n.} Proton imaging snapshots in the reverse configuration (sketched in the frame \textbf{p.}) and typical image of the unperturbed proton beam (\textbf{o.}). In all images, the spatial scale refers to the interaction plane and time is relative to the
arrival of the peak of the CPA$_1$ pulse on target.} \label{raw}
\end{center}
\end{figure*}

The experiment was carried out at the Vulcan laser system in the Rutherford Appleton Laboratory \cite{TAW} using two laser beams both with a central wavelength $\lambda_L=1.05\mu$m, energy $E_L=50$J and duration $\tau_L=1$ps. Both beams were preceded by a lower intensity plateau ($I_P\approx 10^{12}$W/cm$^2$, duration of $\approx$ 300 ps), due to amplified spontaneous emission.
The first laser beam (CPA$_1$) was focussed, down to a focal spot with radius $r_L\approx10\mu$m, to a peak intensity of $I_L\approx 10^{19}$W/cm$^2$ (dimensionless intensity $a_0\approx 2.7$) onto a $d=10\mu$m thick aluminum foil, with a 45$^\circ$ angle of incidence. Hydrodynamic simulations \cite{Hyades} indicate that the laser pre-pulse ablates a sub-micron layer of the aluminum foil that expands, in a plasma state, with an exponentially decreasing density profile (scale-length of 2.5 microns) and an electron temperature of $\approx$100 eV. The second laser beam (CPA$_2$) was focussed onto a $20\mu$m thick gold foil to generate, via Target Normal Sheath Acceleration (TNSA) \cite{Snavely}, a proton beam with a Boltzmann-like spectrum (temperature of $T_P=3.0\pm0.2$ MeV and cut-off energy of $E_p\approx 20$MeV, see Figs. \ref{raw}.g and \ref{raw}.o for its spatial distribution). This beam was used as a charged-particle probe \cite{Sarri} and was recorded onto a stack of calibrated RadioChromic Films (RCFs) \cite{Dempsey} giving a point-like projection of the interaction with a geometrical magnification $M \approx (l+L)/l \approx 11$ with $l\approx 3$ mm and $L\approx 3$cm the distances between the Au and Al foils and between the Al foil and the RCF respectively, as sketched in Figs. \ref{raw} h. and \ref{raw} p.

Previous theoretical modeling of high-intensity irradiation of
thin solid targets \cite{Pukhov} indicates
the generation of magnetic fields having a toroidal structure
with azimuthal symmetry and field lines parallel to the target surface.
The intense electric fields generated in the expanding plasma sheath are
almost normal to the target surface and have been directly detected
using a probe proton beam parallel to the surface \cite{Romagnani}.
In the present experiment, in order to maximize the probe protons deflections
due to magnetic fields over those due to electric fields, the propagation axis
of the probe beam was normal to the target surface. In order to ascertain the magnetic nature of the deflecting fields, two configurations were adopted: the proton beam first encountered either the rear, un-irradiated side of the target (\emph{direct} configuration, Fig. \ref{raw}.h) or the front, irradiated side (\emph{reverse} configuration, Fig. \ref{raw}.p). For a given polarity of the magnetic field distribution, the two configurations should induce
opposite deflections.

\begin{figure}[!b]
\begin{center}
\includegraphics[width=0.8\columnwidth]{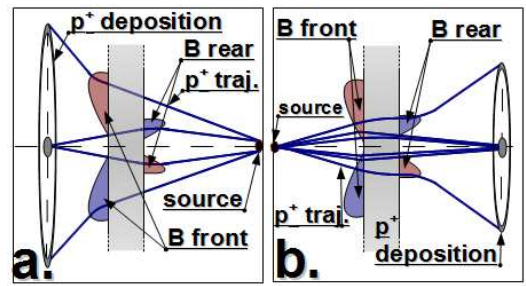}
\caption{Sketch of the proton deflections induced by the magnetic fields in the direct (\textbf{a.}) and reverse (\textbf{b.}) configuration.} \label{sketch}
\end{center}
\end{figure}
A typical set of RCF images, obtained in the direct configuration, are displayed in Figs. \ref{raw} (series a.-f. and i.-k.). All images depict the presence of two main features: an outer ring (radius of the order of 100 - 200 $\mu$m) and an inner dot (radius of the order of 30-40 $\mu$m) of proton accumulation. The outer ring is seen to slowly expand in time whilst roughly preserving its amplitude whereas the inner dot becomes weaker and eventually disappears as time progresses. The reverse configuration (Figs. \ref{raw} l.-n.) induces an inverse deflection pattern, the central dot being much darker than the outer ring.
\begin{figure}[!b]
\begin{center}
\includegraphics[width=0.8\columnwidth]{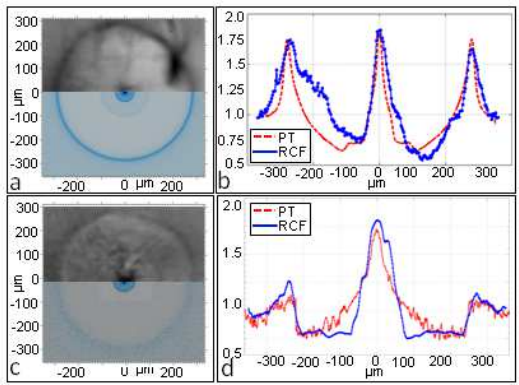}
\caption{Comparison between the experimental proton dose deposition as recorded by the RCF and that resulting from PT simulations assuming magnetic field distributions as the ones depicted in Fig. \ref{sims}. Spatial scales refer to the interaction plane. Frames \textbf{a.} and \textbf{b.} refer to the direct configuration whereas frames \textbf{c.} and \textbf{d.} to the reverse one.}\label{RCF_PT}
\end{center}
\end{figure}
In the direct configuration, the rear magnetic field focusses the protons propagating near the axis whereas the front field enhances the divergence of the protons propagating at a wider angle (see Fig. \ref{sketch}.a). In the radiographs, this leads to the formation of an accumulation dot produced by the rear field and of an outer ring produced by the front field. In the reverse configuration, the same structures should be expected, yet with a different relative amplitude (Fig. \ref{sketch}.b), in qualitative agreement with Fig. \ref{raw}.
\begin{figure}[!t]
\begin{center}
\includegraphics[width=0.9\columnwidth]{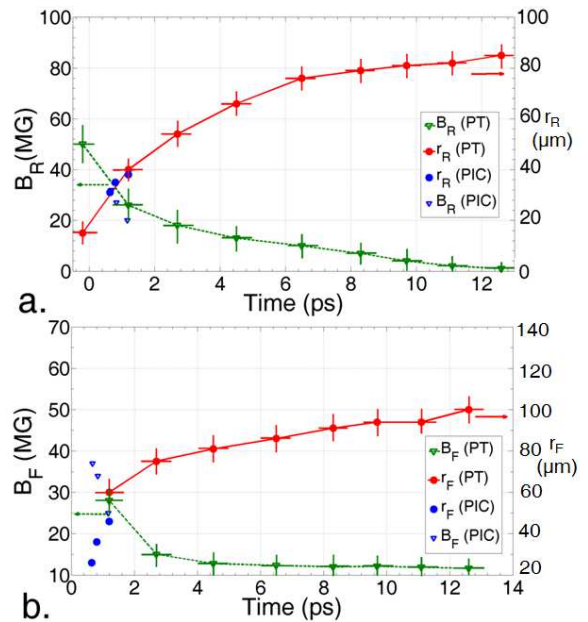}
\caption{Temporal evolution of the magnetic field amplitude and width at the rear (\textbf{a.}) and front (\textbf{b.}) side of the target as extracted from matching PT simulations, compared with the PIC results. Time refers to the arrival of the peak of CPA$_1$ on target.} \label{general}
\end{center}
\end{figure}

For a quantitative analysis of the magnetic field, Particle Tracing (PT) \cite{Romagnanithesis} simulations have been performed. The code follows the propagation of a proton beam, as employed in the experiment, through a prescribed magnetic field distribution and it includes the response of an RCF detector providing synthetic RCF images. The magnetic field is assumed, at both the front and the rear side of the target, as having a cylindrically symmetric toroidal distribution localized at the target surfaces, analogous to the model reported in Ref. \cite{Cecchetti}. In this model, the magnetic field at each of the two surfaces is described via the following parameters: $B$ (field amplitude), $R$ (radius of the field distribution), and $L_r$ and $L_z$ (field scalelength in directions parallel and perpendicular to the target surface, respectively).  These parameters were varied independently until a satisfactory match was reached between experimental and synthetic data, in terms of position and amplitude of the optical density peaks observed on RCF in correspondence to the ring and dot proton accumulation regions (see Fig. \ref{RCF_PT}). PT simulations indicate that variations in $B$ and $R$ at both surfaces affect strongly peak amplitude and radius, respectively, while the field scale-lengths have only a secondary effect on the synthetic RCF profile. By keeping $L_z$ $=$ $L_r$ $=$ 10 $\mu$m (consistent with PIC simulations, see later) and varying independently $B$ and $R$ at front and rear, profile matching as shown in Fig. \ref{RCF_PT} could be obtained, in both the direct and reverse configuration. The matching was considered satisfactory whenever the position and peak value of the simulated optical density maxima reproduced those of the experiment within $\pm$ 5 $\mu$m and 5\%, respectively. These values are of the order of the intrinsic spatial resolution of the proton backlighting \cite{Sarri} and of a typical small-scale non-homogeneity of the proton density on a beam cross section.

By iteratively applying this method to each RCF layer, it has been possible to simultaneously extract the temporal evolution of the amplitude and radius of the magnetic field at each side of the target (see Fig. \ref{general}). In correspondence to the falling edge of the laser pulse, both fields are seen to increase their radius and decrease their amplitude in time. After the laser irradiation ($t \geq 1$ ps in Fig. \ref{general}) the front field is seen to rapidly drop down to an almost constant amplitude of $B_F \approx 10$ - $12$ MG while its radius increases up to an approximately constant value of 90 $\mu$m. On the other hand, the rear field exponentially decreases in amplitude with a typical time-scale of the order of 7 ps. Meanwhile, its radius increases in time with decreasing radial velocity. The longer persistence of magnetic fields at the front surface might be related to the presence of the underdense pre-plasma which is able to better support the magnetic field lines \cite{Frutchman}.

\begin{figure}[!b]
\begin{center}
\includegraphics[width=0.8\columnwidth]{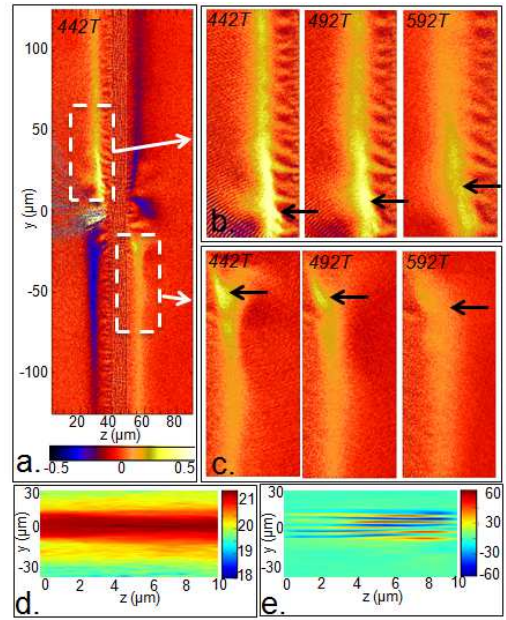}
\caption{\textbf{a.}-\textbf{c.}: PIC simulation results.
The frames show the transverse magnetic field ($B_x$) at $t=442 T$ in units of $B_0=m_e\omega c/e=107$ MG (\textbf{a.}), and a zoom of the front (\textbf{b.}) and rear (\textbf{c.}) regions corresponding to the dashed rectangles in \textbf{a.} at $t=442, 492$ and $592 T$. The black arrows indicate the position of the field maxima.
\textbf{d.}-\textbf{e.}: transport simulation results.
The frames show the hot electron density inside the target in logarithmic scale and units of cm$^{-3}$ (\textbf{d.})
and the related magnetic field distribution in units of MG (\textbf{e.}), both at $t=1$ ps after the peak of the pulse.} \label{sims}
\end{center}
\end{figure}

The presence of MegaGauss magnetic fields of opposite polarity at both
sides of the target, is also supported by
two-dimensional (2D) Particle-In-Cell (PIC) simulations.
A density profile composed by an exponential ramp reproducing the above mentioned preplasma, followed by a plasma bulk with electron density $n_e=40n_c$ and charge-to-mass ratio $Z/A=9/26$ is assumed.
The laser pulse has a Gaussian transverse profile (FWHM = $5~\mu\mbox{m}$), a duration of $250~T$ (with the laser period $T=3.3~\mbox{fs}$ for $1~\mu\mbox{m}$ wavelength), dimensionless peak amplitude $a_0 = 2.7$ on axis and it is incident at $20^\circ$ with respect to the normal of the target surface. The simulation ran up to $600T \simeq 2~\mbox{ps}$, in order to overlap with the proton radiographs at earlier times.
Fig.\ref{sims} shows the generation of magnetic fields at the front and rear surfaces of the target, both having an approximately antisymmetrical distribution with respect to the axis, and polarity opposite to each other. Near the peak of the laser pulse the fields reach a maximum amplitude of the order of 50 MG (Fig. 2.a) and, during the rise of the laser pulse, they propagate in the transverse ($y$) direction with a constant velocity of $\simeq 2.7 \times 10^8~\mbox{m/s}$ (consistently with the scenario experimentally investigated in Ref.\cite{Quinn}) while, as the laser intensity falls down, the field distribution drifts at a much lower velocity of $\sim 2 \times 10^7~$m/s until stagnation is reached ($t \geq 10$ ps). Within the intrinsic approximations that a PIC model unavoidably introduces (such as two-dimensional geometry and non-collisionality) a fair agreement is found between the experimental and numerical results (see Fig. \ref{general}).
The PIC simulations also indicate that the electrostatic fields at the target's surfaces (not shown for brevity) are almost normal to the original surface (i.e. parallel to the main axis of propagation of the probing proton beam), in agreement with reported experimental observations \cite{Romagnani}, and allow to estimate their intensity. The inclusion of such electrostatic fields does not affect significantly the PT images and can thus be neglected.

In principle, strong magnetic fields may also be generated inside the target due to resistive return currents which must balance the fast electron flow.
In order to evaluate these fields, 3D simulations of the propagation of an electron beam through aluminium at an initial temperature of 1~eV were performed, using the code ZEPHYROS \cite{Kar}. The simulation assumed suitable parameters for the electron beam (electron energy of 0.6~MeV, beam density of $4\times10^{20}$cm$^{-3}$, initial radius $r_s=8.5~\mu\mbox{m}$, divergence $\theta_d \simeq 25^{\circ}$ \cite{Fuchs}) and a background-temperature dependent target resistivity \cite{resistivity}. Simulation results show the growth of small-scale filaments \cite{Yuan} with magnetic fields amplitude up to $\simeq 40~\mbox{MG}$ and a characteristic spatial scale of $4~\mu\mbox{m}$ (Fig. 2.e) which is below the proton imaging resolution \cite{Sarri}. Indeed, including magnetic fields in the target bulk with these simulated amplitude and spatial distribution induces in PT images the superposition of random fluctuations with amplitude below 5\%. In these specific experimental conditions, the proton deflections are thus predominantly induced by the magnetic fields generated at the surfaces of the target.

It is of particular interest to analyze in more detail the field dynamics at the rear side of the target, which are much less explored,  and are of  direct relevance to the sheath ion acceleration. Here, the intense magnetic fields are generated by hot electron currents which, when crossing the rear surface, can not be balanced anymore by a counterpropagating return current.
The expected temperature of hot electrons produced in the interaction is $T_h \simeq 0.6~\mbox{MeV}$, corresponding to a velocity $v_h \simeq 0.9c$ and to a relativistic factor $\gamma \simeq 2$; the hot electron density may be thus roughly estimated by a balance of energy fluxes, $f I_L=n_hv_hT_h$ yielding, for an absorption fraction $f\approx0.1$ \cite{Fuchs}, $n_h \simeq 4 \times 10^{20}~\mbox{cm}^{-3}$. The total current due to hot electrons flowing through the target may be estimated as $I_h=en_hv_hS \simeq 5 \times 10^{6}~\mbox{A}$ where $S=\pi r_L^2=3 \times 10^{-6}~\mbox{cm}^{-2}$ is the area of the laser focal spot. The generation of a magnetic field requires a diverging electron flow since, if it were collimated, the large back-holding electric field {\bf E} resulting from charge displacement would cause an equal and anti-parallel displacement current $\textbf{J}_E=\varepsilon_0\partial_t{\bf E}$ of refluxing electrons. In this case the source term for the magnetic field would exactly vanish.
A divergent flow allows part of the current to flow in the radial direction forming loops which fall back to the target where a surface return current may close the circuit. We developed \cite{Andreaarxiv} a simple geometrical model of such ``fountain effect'' to estimate the peak magnetic field as $B_{\mbox{\tiny max}} \simeq \alpha\theta_dB_0$ where $B_0=\mu_0I_h/(2\pi r_0)$, $\theta_d\simeq 25^{\circ}$ is the divergence of the flow, $r_0 \simeq 15~\mu\mbox{m}$ the radius of the electron emitting area and $\alpha \simeq 8T_h/(eE r_0)$ with $E$ the typical value of the electric field. By estimating $E \simeq 10^{12}~\mbox{V m}^{-1}$, as suggested both by the PIC simulations and by proton emission data in similar conditions \cite{Romagnani}, we obtain a peak value of $B_{\mbox{\tiny max}}\simeq 70~\mbox{MG}$, in fair agreement with the experimental results. The corresponding value of the Larmor radius $m_e\gamma v_h/eB \simeq 3B_{10}^{-1}~\mu\mbox{m}$ (where $B_{10}$ is the field in units of 10~MG) is small enough to indicate that electrons would be strongly magnetized in the regions of peak field. In these conditions, the magnetic field lines are frozen in the electron fluid, and the magnetic forces tend to confine the plasma. This tendency may account for the ``negative acceleration'' in the torus radius versus time $d^2r_{\mbox{\tiny F,R}}/dt^2<0$ observed in Fig.\ref{general}. The magnetic energy density ($u_m=B^2/2\mu_0=4\times 10^{5}B_{10}^{2}~\mbox{J cm}^{-3}$) becomes in fact comparable to the maximum plasma thermal energy density expected at the peak of the laser pulse ($u_t=n_hT_h\simeq 4\times 10^{7}~\mbox{J cm}^{-3}$).
Since the typical times for collisional dissipation and magnetic diffusion (of the order of tens of ns)  are much longer than the time scales of the observation, the magnetic field value should mainly decay in time because of the sheath expansion and thus be roughly inversely proportional to the square of the observed radius, in agreement with the observations.

In conclusion, temporally and spatially resolved proton imaging indicates the generation of toroidal magnetic fields having tens of MegaGauss strength on both sides of a foil irradiated by an intense laser pulse. The magnetic fields are strong enough to effectively confine the radial expansion of the plasma region where they are generated thus possibly affecting the ion acceleration in the expanding sheath. Morevoer, the self-confining effect qualitatively resembles the collimation of leptonic astrophysical jets, suggesting that the present framework is suitable to investigate similar mechanisms in down-scaled laboratory experiments.

We acknowledge the support of the RAL-CLF staff. This work has been supported by the Engineering and Physical Sciences  Research  Council  [Grant  Numbers  EP/E035728/1 (LIBRA consortium), EP/D06337X/1 and EP/J002550/1], by British Council-MURST-CRUI, by the Italian Ministry for University and Research (FIRB project ``SULDIS'') and by a Leverhulme Trust fellowship (ECF-2011-383). The PIC simulations were performed using the computing resources granted by the VSR of the Research Center J\"ulich under the project HRO01. L.R. acknowledges support from the ULIMAC grant from the Triangle de la Physique RTRA network.

\end{document}